\documentclass[a4paper, twocolumn]{article}

\usepackage[T1,T2A]{fontenc}
\usepackage[utf8]{inputenc}
\usepackage[english]{babel}
\usepackage[left=1.8cm, right=1.8cm, top=2.5cm, bottom=2cm]{geometry}

\usepackage{url}

\usepackage{abstract}

\usepackage{amssymb}
\usepackage{amsmath}
\usepackage{mathrsfs}
\usepackage{physics}
\usepackage{bm}

\usepackage{cuted}
\usepackage{flushend}
\usepackage{pgfplots}
\pgfplotsset{compat=1.9}
\usepackage{graphicx}
\usepackage{multirow}
\usepackage{float}
\usepackage{subfigure}
\usepackage{wrapfig}
\usepgfplotslibrary{colormaps}

\usepackage{tikz}
\tikzset{>=stealth}
\usetikzlibrary{calc}
\usetikzlibrary{angles, quotes}
\usetikzlibrary{math}
\usepackage{ifthen}

\title{
Logarithmically complex rigorous Fourier space solution\\to the 1D grating diffraction problem
}
\date{}
\author{$^1$Evgeniy Levdik and $^2$Alexey\,A.\;Shcherbakov\\$^1$Univ. degli Studi di Siena, Siena, Italy\\$^2$ITMO University, St.\,Petersburg, Russia}

\begin{document}

\maketitle

\section*{Abstract}
The rigorous solution to the grating diffraction problem is a cornerstone step in many scientific fields and industrial applications ranging from the study of the fundamental properties of metasurfaces to the simulation of photolithography masks. Fourier space methods, such as the Fourier Modal Method, are established tools for the analysis of the electromagnetic properties of periodic structures, but are too computationally demanding to be directly applied to large and multiscale optical structures. This work focuses on pushing the limits of rigorous computations of periodic electromagnetic structures by adapting a powerful tensor compression technique called the Tensor Train decomposition. We have found that the millions and billions of numbers produced by standard discretization schemes are inherently excessive for storing the information about diffraction problems required for computations with a given accuracy, and we show how to adapt the TT algorithms to have a logarithmically growing amount of information to be sufficient for reliable rigorous solution of the Maxwell's equations on an example of large period multiscale 1D grating structures.

\bigskip

\section{Introduction}
The problem of simulating the diffraction of electromagnetic waves by wavelength-scale gratings can be solved efficiently by many methods \cite{antonakakis:hal-00785737}. However, when periodic structures have a multi-scale pattern, conventional rigorous numerical methods rapidly lose their efficiency as the largest scale and the difference between scales increase. As an example, consider a simple 2D structure with two characteristic scales $\Lambda$ and $\Lambda_1$, as shown in Fig.~\ref{fig:grating}, where one has to finely resolve the fields in the region of size $\Lambda_1$, which makes the total number of mesh nodes of length $\Lambda$ quite large, provided that $\Lambda \gg \Lambda_1$. This is the case for metasurfaces, which typically represent 1D/2D periodic structures with complex unit cells and unique scattering properties \cite{Schulz2024}. As they move towards industrial applications, there is a growing need for the most appropriate and powerful simulation techniques.

There are several ways to address the problem of excessive complexity of rigorous electromagnetic computations. The first is to develop hybrid and domain decomposition methods \cite{Bonnet2018,Hudelist2009,Hughes2021,Salary2016,Gao2024}. The second makes intensive use of modern machine learning (ML) methods \cite{10237050,Majorel2022,An2020,Skarda2022}, with physics-informed neural networks in particular being among the most promising \cite{10221390,Sarkar2023}. Finally, there is a space in the implementation of the known methods taking into account the modern mathematical approaches, and the current work fits in this direction.

Fourier space methods such as the Fourier Modal Method (FMM), the Differential Method (DM), the C-Method or the Generalized Source Method (GSM) are among the most powerful for planar grating diffraction problems \cite{Petit1980,SHCHERBAKOV-2012-2DGSM,Chandezon2024}. Among them, the GSM provides solutions with the lowest asymptotic complexity with respect to the Fourier space grid size $N_F$, namely $O(N_F\log(N_F))$. In essence, the GSM can be regarded as a Fourier space volume integral method. The fields are expressed as a superposition of plane waves. The volume integral equation, which provides a solution to the vector Helmholtz equation, can be reduced to a linear system of equations with an internal structure in the form of a product of block diagonal and block Toeplitz matrices \cite{SHCHERBAKOV-2012-2DGSM, SHCHERBAKOV-2010-1DGSM}. Although this method has demonstrated its efficiency for complex unit cells \cite{Shcherbakov2017}, it still requires too many computational resources for scales $\Lambda \gg \lambda$, where $\lambda$ is the wavelength.

\begin{figure}[!htb]
    \centering
    \begin{tikzpicture}
        \def\l{0.5};
        \def\n{4};
        \def\N{3};
        \def\thickness{0.7};
    
        \foreach \i in {1, ..., \N}{
            \foreach \x in {1, ..., \n}{
                \def\mid{(\x - 0.5) * \l + \n * \l * (\i- 1)};
                \def\w{\l * tanh((1 - \x / \n) + 0.2)}
                \fill ({\mid - \w / 2}, 0) rectangle ++({\w}, {-\thickness});
            }
        }
    
        \draw (0, 0) -- ++({\N * \n * \l}, 0) coordinate (a);
        \draw (0, {-\thickness}) -- ++({\N * \n * \l}, 0) coordinate (b);
        \draw[dotted] (0, 0) -- ++(-0.3, 0);
        \draw[dotted] (a) -- ++(0.3, 0);
        \draw[dotted] (0, {-\thickness}) -- ++(-0.3, 0);
        \draw[dotted] (b) -- ++(0.3, 0);
        
        \draw[->, thick, line cap=round] (-0.8, 0) coordinate(o) -- ++(0.5, 0) node[above] {$\scriptstyle x$};
        \draw[->, thick, line cap=round] (o) -- ++(0, -0.5) node[below] {$\scriptstyle z$};
        \draw[<->] ($(a) + (0.1, 0)$) -- ($(b) + (0.1, 0)$) node[midway, right] {$\scriptstyle H$};
    
        \foreach \x in {2, ..., \n}{
            \def\mid{(\x + 0.5) * \l};
            \draw[dashed] ({\mid}, {-\thickness}) -- ++(0, -0.3) coordinate (m);
            \draw[very thin, gray, <->] ($(m) + (0, 0.1)$) -- ++(-\l, 0);
        }
        \draw[semithick, dashed] ({\l * 0.5}, {-\thickness}) -- ++(0, -0.3);
        \draw[dashed] ({\l * 1.5}, {-\thickness}) -- ++(0, -0.3) coordinate (b);
        \draw[semithick, dashed] ({\l * (\n + 0.5)}, {-\thickness}) -- ++(0, -0.3) coordinate (a);
        \draw[semithick, dashed] ({\l * (2 * \n + 0.5)}, {-\thickness}) -- ++(0, -0.3);
        \draw[<->] ($(a) + (0, 0.1)$) -- ++({\l * \n}, 0) node[below, midway] {$\scriptstyle\Lambda$};
        \draw[<->] ($(b) + (0, 0.1)$) -- ++({-\l}, 0) node[below, midway] {$\scriptstyle\Lambda_1$};
    
        \def\m{6};
        \def\vecl{0.7};
        \def\a{30};
        \foreach \i in {1, ..., \m}{
            \draw[thick, ->] ({(\l * \n * \N / \m) * (\i - 0.5) - \vecl * sin(\a) / 2}, {\vecl + 0.5}) coordinate (p) -- ++({\vecl * sin(\a)}, {-\vecl * cos(\a)}) coordinate (o);
        };
        \draw[dashed] ($(o) - (0, 0.2)$) -- ++(0, {\vecl * cos(\a) + 0.35}) coordinate (q);
        \pic [draw,
          angle radius=5mm, angle eccentricity=1.3,
          "$\scriptstyle\theta$"] {angle=q--o--p};
        \node at ({\N * \n * \l / 2}, {0.5 + \vecl / 2}) {$\vb*{k}_0$};
    
        \begin{scope}[shift={(0, -2.5)}]
            \def\low{0.2};
            \def\h{0.5}
            \draw[thick, ->] (0, 0) -- (0, 1) node[left] {$\scriptstyle \varepsilon(x)$};
            \draw[thick, ->] (0, 0) -- ({\n * \l * \N + 0.5}, 0) node[above] {$\scriptstyle x$};
            \draw (0, \low) -- ++(-0.05, 0) node[left] {$\scriptstyle \varepsilon_b$};
            \foreach \i in {1, ..., \N}{
                \foreach \x in {1, ..., \n}{
                    \def\mid{(\x - 0.5) * \l + \n * \l * (\i- 1)};
                    \def\w{\l * tanh((1 - \x / \n) + 0.2)}
                    \draw ({\mid - \l / 2}, \low) -- ({\mid - \w / 2}, \low) -- ++(0, \h) -- ++({\w}, 0) -- ++(0, -\h) -- ({\mid + \l / 2}, \low);
                }
            }
            \draw[dotted] ({\n * \l * \N}, \low) -- ++(0.3, 0);
        \end{scope}
    \end{tikzpicture}
    \caption{Plane wave diffraction on 1D grating with multi-scale pattern. $\Lambda$~-- grating period, $\Lambda_1$~-- local characteristic scale. $\varepsilon(x,y,z) = \varepsilon(x,z)$~-- grating dielectric permittivity spacial distribution periodic in $x$ direction, $\bm{k}_0$~-- incident wave vector, $H$~-- grating depth, $\theta$~-- angle of incidence.}
    \label{fig:grating}
\end{figure}

Since the GSM diffraction matrix has a regular structure, the corresponding calculations can be speeded up dramatically for large matrix sizes. For this purpose, we propose to use the Tensor Train (TT) decomposition \cite{OSELEDETS-2011-TT}. This is a low-rank tensor representation that allows storing vectors and matrices of size $N$ with $O(\log N)$ elements. As well as being a way of compressing data, the TT format supports most tensor algebra operations, such as matrix and matrix-by-vector products. Linear systems of equations can be solved in TT format using the AMEn algorithm \cite{OSELEDETS-2011-TT, DOLGOV-2014-AMEN}. So it's possible to solve an astronomically large linear system without ever calculating a majority of its coefficients.

In this work we aim to investigate how an application of Tensor Train Decomposition within the GSM can speed up practical diffraction calculations. As a proof of concept, we focus on modelling diffraction by 1D gratings with multi-scale patterns. First, we describe an analytical model and review the equations to be solved numerically. Secondly, we introduce the tensor train format, relevant algorithms and describe how it has been applied within the GSM. Finally, we present the results of numerical simulations and evaluate the capabilities of the new method.

\section{The Generalized Source Method}

The detailed derivation of the GSM in a general form can be found in \cite{SHCHERBAKOV-2010-1DGSM, SHCHERBAKOV-2012-2DGSM, TISHCHENKO-2000-GSM}. Here we consider a simple 1D grating in vacuum. The diffraction problem is shown schematically in Fig.~\ref{fig:grating}. The collinear monochromatic diffraction of TE polarized plane waves is considered for gratings of depth $H$ with period $\Lambda$, and the period is assumed to be "pixelated" with pixel size $\Lambda_1$ and filling factor varying between pixels. The Cartesian coordinates are chosen so that the periodicity direction is along the axis $X$ and the axis $Z$ is perpendicular to the grating plane. The periodic permittivity depends only on the $x$ coordinate, $\varepsilon=\varepsilon(x)$, so the grating is a 1D photonic crystal slab.

The solution to the vector Helmholtz equation for the electric field is written as a volume integral equation with the free space Green's dyadic \cite{chew1995waves}. In the simplest case of the TE polarization, this solution takes the following form:

\begin{strip}
\begin{equation}
    E_y(x,z) = E_y^{inc}(x,z) - \frac{\omega\mu_0}{4\pi^2} \int\limits_{-\infty}^{\infty} \dd\zeta \int\limits_{-\infty}^{\infty} \dd k_x\, e^{i k_x x} \, \frac{e^{ik_z|z-\zeta|}}{2k_z} \int\limits_{-\infty}^{\infty} \dd x' \, J_y(x',\zeta)e^{-ik_xx'}.
    \label{eq:inteqEy}
\end{equation}
\end{strip}
Here the wavevector projections $k_{x,z}$ are related via the dispersion equation $k_z = \sqrt{k_0^2-k_x^2}$, $k_0^2=\omega^2\varepsilon_0\mu_0$.

To obtain a self-consistent equation the electric current is replaced by a generalized source $J_y^\text{gen} = -i\omega(\varepsilon(x)-\varepsilon_0)E_y$, which exists within the grating layer only, $-H/2\leq z\leq H/2$. Then the Floquet-Bloch theorem is invoked and the Fourier decomposition of all periodic functions is applied. The field amplitude is replaced by a superposition of plane wave amplitudes propagating up and down with respect to the axis $Z$. Then, Eq.~(\ref{eq:inteqEy}) for a fixed Bloch wavevector $k_{x0}$ becomes
\begin{align}
    a_m^+(z) &= a_m^{\text{inc},+}(z) + \frac{ik_0^2}{2k_{zn}} \int\limits_{-H/2}^{z} e^{ik_z(z-\zeta)} J_{my}^\text{gen}(\zeta) \, \dd \zeta \\
    a_m^-(z) &= a_m^{\text{inc},-}(z) + \frac{ik_0^2}{2k_{zn}} \int\limits_{z}^{H/2} e^{ik_z(\zeta-z)} J_{my}^\text{gen}(\zeta) \, \dd \zeta
    \label{eq:inteq_amp}
\end{align}
with the grating equation $k_{xm}=k_{x0}+2\pi m/\Lambda$, $m\in\mathbb{Z}$. $J_{my}^\text{gen}$ denote Fourier amplitudes of the generalized current.

Upon the Fourier transform the product $\Delta\varepsilon E_y$ becomes a convolution product. Discretization of the integrals using the mid-point rule with the step $\Delta z=h$ and the total number of slices along $Z$ direction $N_S$ gives an infinite system of linear equations
\begin{equation}
    a_{mk}^{\pm} = a_{mk}^{\text{inc},\pm} + \frac{ik_0}{2k_{zn}} \sum\limits_{q=1}^{N_S} \Theta^{\pm}_{m,kq} \sum_n \left(\frac{\Delta\varepsilon}{\varepsilon_0}\right)_{\!m-n} \!\! (a_{nq}^+ + a_{nq}^-)
    \label{eq:algsys}
\end{equation}
where 
\begin{equation}
\begin{split}
    \Theta^{\pm}_{m,kq} = \frac{k_0 h}{2}\left[
(1-\delta_{kq})e^{\pm ik_{zm}h(k-q)} \right. + \\
 \left. + \delta_{kq}e^{ik_{zm}h/2} \right] \theta\left(\pm(k-q)\right)
    \label{eq:theta}
\end{split}
\end{equation}
is written via the Kronecker $\delta$-symbol and the Heaviside step function $\theta$. Truncation of the infinite Fourier series, $|n|\leq\lfloor N_F/2 \rfloor$, gives a finite system of linear algebraic equations, which we write though a specific matrix products:
\begin{equation}
    \bm{a} = \bm{a}^\text{inc} + \bm{P}_b\bm{Y}\bm{D}\bm{X}\bm{a}
    \label{eq:linear-system}
\end{equation}
The size of the system is $2 N_F N_S$. The block-diagonal matrices $\bm{X}$ and $\bm{Y}$ denote the forward and backward conversion from the plane wave amplitudes to the Fourier amplitudes of the electric field projection $E_{ym}$. The block-Toeplitz matrix $\bm{D}$, which elements are proportional to the Fourier amplitudes of the permittivity function, can be interpreted as an emission operator in each slice by the generalized sources. Finally, the block diagonal matrix $\bm{P}_b$ describes the propagation of the emitted plane waves between the slices within the grating layer. Explicitly the matrix elements are specified further.

Once the Eq.~(\ref{eq:linear-system}) is solved, its solution is used to find amplitudes of diffracted waves propagating away from the grating layer boundaries $z=\pm H/2$ by invoking the Eq.~(\ref{eq:algsys}) one more time:
\begin{equation}
    \bm{a}^\text{out} = \bm{a}^\text{inc} + \bm{TYDX}\bm{a}
    \label{eq:output}
\end{equation}
Here the matrix $\bm{T}$ physically corresponds to propagation of the emitted waves from each slice to the grating layer boundaries.

Asymptotic complexity of the GSM is $O(n_{it}N_FN_S\log (N_FN_S))$, which is governed by the Fast Fourier Transform which is used for fast multiplication of vectors by Toeplitz matrices \cite{SHCHERBAKOV-2012-2DGSM}. Number of iterations of a linear system solver $n_{it}$ (e.g., the Generalized Minimal Residual method or Biconjugate Gradient Stabilized method) strongly depends on the grating geometry and composition. For a given grating structure with sharp material interfaces the method has polynomial convergence with respect to both $N_F$ and $N_L$.

\section{Tensor Train Decomposition}
Tensor Train (TT) Decomposition is a method for storing tensors, including vectors and matrices, in a compact way. The TT format supports all basic tensor operations (such as summation, transposition, and element-wise products, etc.) allowing results to be computed directly in the TT format without the need for decompression \cite{OSELEDETS-2011-TT}. This format, along with various related algorithms, is implemented in the Python library \texttt{ttpy} by Ivan Oseledets \cite{ttpy}, which we used for our calculations. For consistency, we provide a brief outline of the key definitions and algorithms essential for our simulations.

\subsection{Definition and basic properties}
A $d$-dimensional tensor $T$ of size $n_1 \times \ldots \times n_d$ is said to be in TT format if there exist $d$ 3D tensors $\mathcal{T}_k$ of size $r_{k - 1} \times n_k \times r_k$, where $k = 1\ldots d$, such that:
\begin{equation}
    T_{i_1 i_2 \ldots i_d} = \mathcal{T}_1(i_1) \cdot \mathcal{T}_2(i_2) \cdot \ldots \cdot \mathcal{T}_d(i_d)
    \label{equation:tt-def}
\end{equation}
3D tensors $\mathcal{T}_k$ are called \textit{cores}, and their sizes $r_k$ are called \textit{ranks} where $r_0 = r_d = 1$. $\mathcal{T}_k(i_k)$ are 2D slices of cores and <<$\cdot$>> stands for matrix multiplication. Since $r_0 = r_d = 1$, the first slice is a row vector and the last slice is a column vector, thus the resulting product is a scalar. This definition is visualized in Fig.~\ref{fig:tt-def}.

\begin{figure}[!htb]
    \centering
    \def\dx{0.15}
    \def\dy{0.2}
    \def\a{0.3}
    \def\op{0.8}
    \definecolor{clr1}{RGB}{0,114,178}
    
    \def\ttcore#1#2#3#4#5{  
    \begin{scope}[shift={#1}]
        \foreach \i in {1, ..., #2}{
            \tikzmath{\n=(#2-#5+1);}
            \ifthenelse{\i=\n}{\def\slicecolor{clr1!50}}{\def\slicecolor{white}}
            \def\sh{(-\i + 0.5 + #2/2)}
            \fill[\slicecolor, opacity=\op] ({\sh*\dx - #4*\a/2}, {\sh*\dy - #3*\a/2}) rectangle ++({#4*\a}, {#3*\a});
            \draw[opacity=\op, step=\a, shift={({\sh*\dx - #4*\a/2}, {\sh*\dy - #3*\a/2})}] (0, 0) grid ({#4*\a}, {#3*\a});
        }
    \end{scope}
    }
    
    \begin{tikzpicture}[scale=0.9]
        \node[left] at (0.2, 0) {$T_{2102} =$};
        \ttcore{(1, 0)}{4}{1}{4}{3};
        \node at (2, 0) {$\times$};
        \ttcore{(3.2, 0)}{3}{4}{5}{2};
        \node at (4.4, 0) {$\times$};
        \ttcore{(5.2, 0)}{2}{5}{3}{1};
        \node at (6, 0) {$\times$};
        \ttcore{(6.6, 0)}{4}{3}{1}{3};
    
        \node at (1, -1) {$\mathcal{T}_1(2)$};
        \node at (3.2, -1.3) {$\mathcal{T}_2(1)$};
        \node at (5.2, -1.3) {$\mathcal{T}_{3}(0)$};
        \node at (6.6, -1.3) {$\mathcal{T}_{4}(2)$};
    \end{tikzpicture}
    \caption{Illustration of the TT decomposition. Highlighted matrices are being multiplied to get an element of a 4D tensor of size $4 \times 3 \times 2 \times 4$. Indexing starts from zero.}
    \label{fig:tt-def}
\end{figure}

If $r = \max(r_k)$ and $n = \max(n_k)$, the TT representation of a tensor contains $O(dr^2n)$ elements, compared to the $N = O(n^d)$ elements in the original tensor. Importantly, the ranks $r_k$ are not directly determined by the tensor size but rather reflect the compression rate. While the TT representation of an arbitrary tensor can sometimes result in large ranks, potentially exceeding the size of the uncompressed tensor, a regularly structured tensor often admits a representation with much smaller ranks. Moreover, when an exact decomposition is not required, allowing for a low-rank approximation can make the TT representation significantly more compact. If $r$ remains constant for any $d$, the size of the TT-decomposed tensor scales proportionally to $d$, or equivalently $d \sim \log(N)$.

\subsection{Matrices and vectors in TT format}
The higher the dimensionality of a tensor, the more efficient its compression in TT format. However, most tensors encountered in numerical analysis are 2D (matrices) or 1D (vectors). To leverage the logarithmic compression benefits of the TT format, these matrices and vectors are first reshaped into multidimensional tensors before decomposition \cite{OSELEDETS-2011-TT}.

A vector $\bm{x}$ of size $N = n_1 \cdot \ldots \cdot n_d$ is said to be in TT format if it is treated as a $d$-dimensional tensor $X_{i_1 \ldots i_d}$ with mode sizes $n_k$, and this tensor is in the TT format. For simplicity, multi-index notation is used, i.e. the indices of a tensor can be treated as a \textit{long} index and vice versa. $i_1$ is the slowest changing outer index, and $i_d$ is the fastest changing inner index. The relation between $i$ and $i_k$ is as follows:
\begin{equation}
    i = i_1 N_{d-1} + i_2 N_{d-2} + \ldots + i_{d-1} N_1 + i_d,
\label{equation:multiindex-def}
\end{equation}
where $N_k = n_1 \cdot \ldots \cdot n_k$, and the indexing starts at zero. The reverse calculation can be done iteratively with integer division and modulus.

Matrices in the TT format are slightly more complicated, which is motivated by a need for a matrix-by-vector product algorithm that can operate with TT formats without expanding tensors to a full size. A matrix $\bm{A}$ of size $N \times M$ is said to be in TT format if $N = n_1 \cdot \ldots \cdot n_d$, $M = m_1 \cdot \ldots \cdot m_d$, and:
\begin{equation}
    \bm{A}_{ij} = A_{i_1 \ldots i_d,\, j_1 \ldots j_d} = \mathcal{A}_1(i_1 j_1) \cdot \ldots \cdot \mathcal{A}_d(i_d j_d),
\end{equation}
where either $i_k j_k$ is treated as a long index or the cores $\mathcal{A}_k$ are treated as 4D tensors.
This definition implies that $N$ and $M$ can be factorized with the same number of factors $d$. If this is not the case (for non-square matrices), it is still possible to define TT format by simply setting the excessive $n_k$ or $m_k$ to 1.

For matrices and vectors reshaped into multidimensional tensors having corresponding shapes, there were developed algorithms to perform matrix-by-vector and matrix-by-matrix multiplications \cite{OSELEDETS-2011-TT}, which are essential for solving the linear system~(\ref{eq:linear-system}).
The solution can be found by conventional iterative methods, where tensor algebra operations are performed on TT-formatted objects instead of full-size ones. In this paper, the AMEn algorithm was used, which is specifically optimized for the TT format \cite{DOLGOV-2014-AMEN}.

Another important fact about matrix multiplication in TT format is that it results in a TT matrix with ranks equal to the sum of the ranks of the arguments. To prevent rank growth, TT rounding is used after each multiplication \cite{OSELEDETS-2011-TT}, which is an algorithm for finding the lowest rank approximation with a fixed accuracy.

The most common way of compressing arbitrary vectors is so called QTT (quantized tensor train) format, where all $n_k = 2$. In this work, QTT is used where possible.

\subsection{Notation summary}
In this article the following notation is used:
\begin{center}
    \begin{tabular}{c|c|c}
         & \textbf{Vector} & \textbf{Matrix} \\ \hline
        Object & $\bm{x}$ & $\bm{A}$ \\
        Element & $\bm{x}_i$ & $\bm{A}_{ij}$ \\
        TT form & $X$ & $A$ \\
        TT cores & $\mathcal{X}_k$ & $\mathcal{A}_k$ \\
        Core <<slice>> & $\mathcal{X}_k(i_k)$ & $\mathcal{A}_k(i_k, j_k)$
    \end{tabular}
\end{center}

Each object, its TT representation and cores are denoted by the same letter in different fonts. Matrix elements may also be identified by lower case letters if the meaning is obvious from the context.

\subsection{TT-Cross}
One of the main reasons for the widespread adoption of the Tensor Train decomposition is its ease of computation for arbitrary tensors. When a tensor is fully stored in memory, the TT decomposition can be obtained using the TT-SVD algorithm, which generalizes the singular value decomposition (SVD) to tensors of higher dimensions \cite{OSELEDETS-2011-TT}. However, this method becomes impractical for extremely large tensors, as storing all their elements is often infeasible. To address this limitation, the TT framework includes the TT-Cross approximation algorithm, which constructs TT cores by evaluating only a small subset of tensor elements, bypassing the need for full tensor calculation.

It should be noted that the TT-Cross cannot work with arbitrary tensors, since it is impossible to obtain information about a tensor element without calculating it. However, the method is highly effective for \textit{smooth} tensors, which are defined by sampling smooth continuous multivariable functions.

The key idea behind this algorithm is the skeleton decomposition of matrices. If a matrix $\bm{M}$ has rank $r$, it can be recovered from its $r$ linearly independent rows that form a submatrix $\bm{I}$ and columns forming a submatrix $\bm{J}$. If $\bm{A}$ is the matrix formed by intersections of selected rows and columns \cite{Gantmacher-1959-Matrices}:
\begin{equation}
    \bm{M} = \bm{I} \cdot \bm{A}^{-1} \cdot \bm{J}
\end{equation}
Skeleton decomposition works well with low-rank matrices only, but a high-rank matrix can be approximated with a low-rank one. Considering a fixed rank $r$, the best approximation has the largest $\det(A)$ \cite{Goreinov-1997-pseudo-skeleton}.

The TT-Cross algorithm generalizes the skeleton approximation to the multidimensional case. If any element of a $d$-dimensional tensor of size $N = n^d$ can be computed in $O(1)$ operations, then the overall complexity of finding its TT decomposition with ranks $r$ is $O(d r^2 n)$, which is proportional to $\log(N)$ \cite{OSELEDETS2010TTCROSS}.

\subsection{Kronecker product and tiled arrays} \label{ssec:kron}
As mentioned earlier, all basic linear algebra operations can be carried out by directly manipulating TT cores, without expanding the tensors to their full size. Among these operations, the Kronecker product is particularly important for constructing the scattering matrix in TT format. If a matrix $\bm{A}$ is of size $n \times m$ and a matrix $\bm{B}$ is of size $p \times q$, their Kronecker product is a block matrix of size $np \times mq$:
\begin{equation}
    \bm{C} = \bm{A} \otimes \bm{B} = 
    \begin{bmatrix}
        a_{11} \bm{B} & \ldots & a_{1m} \bm{B} \\
        \vdots & \ddots & \vdots \\
        a_{n1} \bm{B} & \ldots & a_{nm} \bm{B}
    \end{bmatrix}
\end{equation}
This operation in the TT format can be computed by simply concatenating the cores of two matrices. If $\bm{A}$ and $\bm{B}$ are in TT format with orders $d_1$ and $d_2$ respectively, then their Kronecker product $\bm{C}$ in TT format will have order $d_1 + d_2$ with cores \cite{OSELEDETS-2011-TT}:
\begin{equation}
\mathcal{C}_k = 
\begin{cases}
    \mathcal{A}_k, & k = 1, \ldots, d_1 \\
    \mathcal{B}_k, & k = d_1 + 1, \ldots, d_1 + d_2
\end{cases}
\end{equation}

Vectors can be treated as matrices of size $1 \times N$, so the Kronecker product can be applied to vectors as well. This is useful for creating tiled vectors. If $\bm{x}$ is a $d$-dimensional TT vector of size $N = n_1 \times \ldots \times n_d$, and vector $\bm{y}$ is the same vector repeated $M = m_1 \times \ldots \times m_{d_1}$ times:
\begin{equation*}
    \bm{y} =
    \begin{bmatrix}
        1 \\
        \vdots \\
        1
    \end{bmatrix}
    \otimes \bm{x}
\end{equation*}
which is in TT form:
\begin{equation}
    \mathcal{Y}_k = 
    \begin{cases}
        [1]_{1 \times m_k \times 1} & k = 1, \ldots, d_1 \\
        \mathcal{X}_k & k = d_1 + 1, \ldots, d_1 + d
    \end{cases}
\end{equation}
An example of vector tiling is shown in Fig.~\ref{fig:tiled}.

\begin{figure}[!htb]
    \centering
    \def\a{0.3}
    \def\op{0.8}
    
    \definecolor{clr1}{RGB}{0,114,178}
    
    \def\ttcore#1#2#3#4#5{  
    \def\dx{0.15}
    \def\dy{0.2}
    \begin{scope}[shift={#1}]
        \foreach \i in {1, ..., #2}{
            \tikzmath{\n=(#2-#5+1);}
            \ifthenelse{\i=\n}{\def\slicecolor{blue!50}}{\def\slicecolor{clr1!50}}
            \def\sh{(-\i + 0.5 + #2/2)}
            \fill[\slicecolor, opacity=\op] ({\sh*\dx - #4*\a/2}, {\sh*\dy - #3*\a/2}) rectangle ++({#4*\a}, {#3*\a});
            \draw[opacity=\op, step=\a, shift={({\sh*\dx - #4*\a/2}, {\sh*\dy - #3*\a/2})}] (0, 0) grid ({#4*\a}, {#3*\a});
        }
    \end{scope}
    }
    
    \def\ttcoreone#1#2#3#4#5{  
    \def\dx{0.1}
    \def\dy{0.17}
    \begin{scope}[shift={#1}]
        \foreach \i in {1, ..., #2}{
            \tikzmath{\n=(#2-#5+1);}
            \ifthenelse{\i=\n}{\def\slicecolor{blue!50}}{\def\slicecolor{white}}
            \def\sh{(-\i + 0.5 + #2/2)}
            \fill[\slicecolor, opacity=\op] ({\sh*\dx - #4*\a/2}, {\sh*\dy - #3*\a/2}) rectangle ++({#4*\a}, {#3*\a});
            \draw[opacity=\op, step=\a, shift={({\sh*\dx - #4*\a/2}, {\sh*\dy - #3*\a/2})}] (0, 0) grid ({#4*\a}, {#3*\a});
            \node[opacity=\op] at ({\sh*\dx}, {\sh*\dy}) {$\scriptstyle 1$};
        }
    \end{scope}
    }
    
    \begin{tikzpicture}[scale=0.9]
        \node[left] at (0.2, 0) {$Y=$};
        
        \ttcoreone{(0.4, 0)}{2}{1}{1}{0};
        \node at (0.9, 0) {$\times$};
        \ttcoreone{(1.35, 0)}{2}{1}{1}{0};
        \node at (1.8, 0) {$\times$};
    
        \ttcore{(2.6, 0)}{4}{1}{3}{0};
        \node at (3.5, 0) {$\times$};
        \ttcore{(4.2, 0)}{3}{3}{2}{0};
        \node at (5.3, 0) {$\times \ldots \times$};
        \ttcore{(6.3, 0)}{4}{4}{1}{0};
    
        \node at (2.6, -0.9) {$\mathcal{X}_1$};
        \node at (4.2, -1.1) {$\mathcal{X}_2$};
        \node at (6.3, -1.3) {$\mathcal{X}_{d}$};
    \end{tikzpicture}
    \caption{TT representation $X$ of a vector $\bm{x}$ with cores $\mathcal{X}_i$ tiled 4 times by adding two cores of size $1 \times 2 \times 1$ with all elements equal to 1. The resulting vector $\bm{y}$ has TT representation $Y$.}
    \label{fig:tiled}
\end{figure}

\subsection{Diagonal, block-diagonal and Toeplitz matrices} \label{ssec:diag}
If a vector $\bm{x}$ is in the TT format with $d$ orders, where each mode has size $n_k$, and with corresponding cores $\mathcal{X}_k$, the diagonal matrix $\bm{A}$ with its elements on the main diagonal, $\bm{A}_{ij} = \delta_{ij} \bm{x}_i$, can be constructed in closed form. The matrix cores $\mathcal{A}_k$ will have $n_k^2$ layers and can be defined as:
\begin{equation}
\mathcal{A}_k(i_k j_k) = 
\begin{cases}
    [0]_{r_{k-1} \times r_k}, & i_k \neq j_k \\
    \mathcal{X}_k(i_k), & i_k = j_k
\end{cases}
\end{equation}
Matrix elements are non-zero only if $i_k = j_k$ for all $k = 1, \ldots, d$, which implies that $i = j$. In this case, the corresponding matrix element is equal to $x_{i}$. TT Cores of an example diagonal matrix are shown in Fig.~\ref{fig:diag}.

\begin{figure}[!htb]
    \centering
    \def\dx{0.15}
    \def\dy{0.2}
    \def\a{0.3}
    \def\op{0.8}
    \definecolor{clr1}{RGB}{0,114,178}
    
    \def\ttcore#1#2#3#4#5{  
    \begin{scope}[shift={#1}]
        \foreach \i in {1, ..., #2}{
            \tikzmath{\n=(#2-#5+1);}
            \def\sh{(-\i + 0.5 + #2/2)}
            \def\slicecolor{clr1!50}
            \fill[\slicecolor, opacity=\op] ({\sh*\dx - #4*\a/2}, {\sh*\dy - #3*\a/2}) rectangle ++({#4*\a}, {#3*\a});
            \draw[opacity=\op, step=\a, shift={({\sh*\dx - #4*\a/2}, {\sh*\dy - #3*\a/2})}] (0, 0) grid ({#4*\a}, {#3*\a});
        }
    \end{scope}
    }
    
    \def\ttcorediag#1#2#3#4#5{  
    \begin{scope}[shift={#1}]
        \tikzmath{\n=(#2 * #2 - 1);}
        \foreach \i in {0, ..., \n}{
            \tikzmath{\n=(#2-#5+1);}
                \tikzmath{
                    \ii=(int(\i / #2));
                    \jj=(\i - #2 * \ii);
            }
            \ifthenelse{\ii=\jj}{\def\slicecolor{clr1!50}}{\def\slicecolor{white}}
            \def\sh{(-\i - 0.5 + \n/2)}
            \fill[\slicecolor, opacity=\op] ({\sh*\dx - #4*\a/2}, {\sh*\dy - #3*\a/2}) rectangle ++({#4*\a}, {#3*\a});
            \draw[opacity=\op, step=\a, shift={({\sh*\dx - #4*\a/2}, {\sh*\dy - #3*\a/2})}] (0, 0) grid ({#4*\a}, {#3*\a});
    
            \ifthenelse{\ii=\jj}{}{
                \foreach \zx in {1, ..., #4}{
                    \foreach \zy in {1, ..., #3}{
                        \node[opacity={\op / 2}, gray] at ({\sh*\dx - #4 * \a/2 + (\zx - 0.5) * \a}, {\sh*\dy - #3 * \a/2 + (\zy - 0.5) * \a}) {$\scriptstyle 0$};
                    }
                }
            }
        }
    \end{scope}
    }
    
    \begin{tikzpicture}[scale=0.9]
        \begin{scope}[shift={(0.35, 0)}]
            \node[left] at (0.2, 2.3) {$X=$};
            \ttcore{(0.8, 2.3)}{2}{1}{3}{0};
            \node at (1.7, 2.3) {$\times$};
            \ttcore{(2.4, 2.3)}{3}{3}{2}{0};
            \node at (3.5, 2.3) {$\times \ldots \times$};
            \ttcore{(4.5, 2.3)}{2}{4}{1}{0};
        \end{scope}

        \node[left] at (0.2, 0) {$A=$};
    
        \ttcorediag{(0.8, 0)}{2}{1}{3}{0};
        \node at (1.7, 0) {$\times$};
        \ttcorediag{(2.7, 0)}{3}{3}{2}{0};
        \node at (4.0, 0) {$\times \ldots \times$};
        \ttcorediag{(5.2, 0)}{2}{4}{1}{0};
    \end{tikzpicture}
    \caption{TT representation $X$ of a vector $\bm{x}$ with cores $\mathcal{X}_i$ turned into TT matrix $A$ representing a diagonal matrix $\bm{A}_{ij} = \delta_{ij} \bm{x}_i$ by adding zero-layers into cores where $i_k \neq j_k$.}
    \label{fig:diag}
\end{figure}

Block-diagonal matrices in TT format can be constructed from TT tensors, containing their blocks. If $X$ is a $d_1 + d$ dimensional tensor of size $m_1 \times \ldots \times m_{d_1} \times n_1^2 \times \ldots \times n_d^2$, block diagonal matrix $\bm{A}$ of size $m^2_1 \times \ldots \times m^2_{d_1} \times n_1^2 \times \ldots \times n_d^2$ can be obtained in TT format as follows:
\begin{equation}
    \mathcal{A}_k(i_k j_k) = 
    \begin{cases}
        [0]_{r_{k-1} \times r_k}, & k \leq d_1 \text{ and } i_k \neq j_k \\
        \mathcal{X}_k(i_k), & k \leq d_1 \text{ and } i_k = j_k \\
        \mathcal{X}_k(i_k j_k), & k > d_1
    \end{cases}
\label{equation:block-diag-cores}
\end{equation}

$\bm{A}$ is the matrix with $M = m_1 \times \ldots \times m_{d_1}$ blocks of size $N = n_1 \times \ldots \times n_d$. If any of the first $d_1$ pairs of indices $i_k$ and $j_k$ are not equal, it indicates that the element lies outside the diagonal blocks, and thus, the element is zero. If an element is within a block, i.e., if $i_k = j_k$ for $k = 1, \ldots, d_1$, the matrix element will be the same as in $\bm{X}$.

A Toeplitz matrix $\bm{A}$ of size $N$ is defined by its anti-diagonal vector $\bm{a}$ of size $2N - 1$: $\bm{A}_{ij} = \bm{a}_{i-j}$. If the TT representation of $\bm{a}$ is given, the TT cores of $\bm{A}$ can be found as shown in \cite{Kazeev-2013-TTToeplitz}.

\subsection{Elementary functions in the TT format} \label{ssec:elementary}
A vector of samples from any smooth continuous function $f(x)$ can be represented in the TT format using the TT-Cross algorithm. However, the resulting cores may not always be optimal due to the algorithm's iterative nature and random initialization. For certain elementary functions, such as $\sin(x)$, $\cos(x)$, and $\exp(ix)$, equidistant samples can be derived analytically, resulting in TT representations with small ranks. These functions are implemented in the \texttt{ttpy} library, although to our knowledge, their formal descriptions have not been published.

Consider the vector $\bm{x}_n = \sin(\alpha n + \varphi)$, where $n = 0, 1, \ldots, N$. If $N = n_1 \cdot \ldots \cdot n_d$, this vector can be represented as a $d$-dimensional TT vector. To achieve this, we will use the following matrix identities:
\begin{equation}
\begin{split}
    &\begin{bmatrix}
        \cos(nx) & \sin(nx)
    \end{bmatrix}
    \cdot
    \begin{bmatrix}
        \cos(mx) & \sin(mx) \\
        -\sin(mx) & \cos(mx)
    \end{bmatrix}
    = \\
    &=\begin{bmatrix}
        \cos((n+m)x) & \sin((n+m)x)
    \end{bmatrix}
\end{split}
\label{equation:sin-cores-horizontal}
\end{equation}
\begin{equation}
\begin{split}
    &\begin{bmatrix}
        \cos(nx) & \sin(nx)
    \end{bmatrix}
    \cdot
    \begin{bmatrix}
        \sin(mx + \varphi) \\
        \cos(mx + \varphi)
    \end{bmatrix}
    = \\
    &= \sin((n+m)x + \varphi)
\end{split}
\label{equation:sin-cores-vertical}
\end{equation}
This allows constructing cores $\mathcal{X}_k$ of vector $\bm{x}$ as:
\begin{equation*}
    \mathcal{X}_1(i_1) = 
    \begin{bmatrix}
        \cos(i_1 \alpha N_{d - 1}) & \sin(i_1 \alpha N_{d - 1}) \\
    \end{bmatrix}
\end{equation*}
\begin{equation}
\begin{split}
    &\mathcal{X}_k(i_k) =
    \begin{bmatrix}
        \cos(i_k \alpha N_{d - k}) & \sin(i_k \alpha N_{d - k}) \\
        -\sin(i_k \alpha N_{d - k}) & \cos(i_k \alpha N_{d - k})
    \end{bmatrix}, \\
    &k = 2, \ldots, d-1
\end{split}
\label{equation:sin-cores}
\end{equation}
\begin{equation*}
    \mathcal{X}_d(i_d) =
    \begin{bmatrix}
        \sin(i_d \alpha + \varphi) \\
        \cos(i_d \alpha + \varphi)
    \end{bmatrix}
\end{equation*}
where $N_k = n_1 \cdot n_2 \cdot \ldots \cdot n_k$. TT ranks of this vector are $r_k = 2$. Replacing (\ref{equation:sin-cores}) with (\ref{equation:sin-cores-horizontal}) and (\ref{equation:sin-cores-vertical}) gives:
\begin{equation}
    \bm{x}_n = \sin((i_1 N_{d-1} + \ldots + i_d) \alpha + \varphi)
\end{equation}
which is the desired $\sin(\alpha n + \varphi)$ according to Eq.~(\ref{equation:multiindex-def}).

The cosine function can be easily constructed using the identity $\cos(\alpha n + \varphi) = \sin(\alpha n + \varphi + \pi / 2)$, and the exponential function $\exp(\alpha n + \varphi)$ can be represented using Euler's formula.

\section{Diffraction problem in the TT format}

To solve the linear system~(\ref{eq:linear-system}) using the AMEn algorithm, we first need to obtain TT representations of both the matrix $\bm{A} = \bm{I} - \bm{P}_b \bm{Y}\bm{D}\bm{X}$ and the vector $\bm{a}^\text{inc}$. After solving the linear system, the diffracted field can be determined by computing the matrix $\bm{B} = \bm{T}\bm{Y}\bm{D}\bm{X}$ (see Eq.~(\ref{eq:output})). While the TT-Cross can be used to approximate participating matrices, a direct evaluation of $\bm{A}_{ij}$ or $\bm{B}_{ij}$ requires a computational cost proportional to their full size. To maintain logarithmic complexity, each matrix ($\bm{I}$, $\bm{P}_b$, $\bm{Y}$, $\bm{D}$, $\bm{X}$, and $\bm{T}$) have to be be decomposed individually into the TT format and then multiplied. To prevent rank growth, TT rounding should be applied after each multiplication.

As mentioned above, the TT-Cross algorithm only works well with smooth tensors. Although the tensors listed above are regularly structured, they are defined by discontinuous functions. For example, a direct application of the TT-Cross to the diagonal matrix $\bm{Y}$ would most likely result in a zero matrix decomposition, as the majority of its elements are zeros. To find compact TT representations, individual structural features of the vectors and matrices have been exploited.

The QTT format was used, so all the vector cores have two layers, and the matrix cores have four layers. All the matrices, except $\bm{T}$, are square with a size of $2 N_F N_S = 2^{d_F + d_S + 1}$, where $N_F = 2^{d_F}$ is the number of Fourier harmonics considered, and $N_S = 2^{d_S}$ is the number of grating slices. The vectors of the diffraction order amplitudes $\bm{a}^{\pm}_{n}(z_p)$ are arranged as follows: $n$ is the inner fastest-changing index, $p$ is the middle index, and $\pm$ is the outer slowest-changing index.

The identity matrix $\bm{I}$ in the QTT format has all ranks $r_k = 1$. Its cores contain four elements: $\mathcal{I}_k(0, 0) = \mathcal{I}_k(1, 1) = 1$ and $\mathcal{I}_k(0, 1) = \mathcal{I}_k(1, 0) = 0$. This ensures that the matrix element is 1 when the row and column indices are equal and 0 otherwise.

\subsection{Amplitude vectors}

Let us denote multiindices:
\begin{equation}  
    \bm{a}^{\pm}_{n}(z_p) = A_{\pm \, p_1 \ldots p_{d_S} \, n_1 \ldots n_{d_F}}  
\end{equation}  
where $A$ is the TT representation of $\bm{a}$, $\overline{n_1 \ldots n_{d_F}}$ and $\overline{p_1 \ldots p_{d_S}}$ are $n$ and $p$ in binary, respectively. With this indexing, the first core of all amplitude vectors in the TT format will be called the \textit{direction core}, with the first layer being selected for $\bm{a}^{+}$ and the second layer for $\bm{a}^{-}$. The next $d_S$ cores will be called \textit{slice cores}, and the last $d_F$ cores will be called \textit{order cores}:  
\begin{equation}  
\begin{split}  
    \bm{a}^{\pm}_{n}(z_p) = &\underbrace{\mathcal{A}_1(\pm)}_{\substack{\text{direction} \\ \text{core}}}  
    \underbrace{\mathcal{A}_2(p_1) \ldots \mathcal{A}_{d_S+1}(p_{d_S})}_{\substack{\text{slice} \\ \text{cores}}} \cdot \\  
    \cdot &\underbrace{\mathcal{A}_{d_S+2}(n_{1}) \ldots \mathcal{A}_{d_S+d_F+1}(n_{d_F})}_{\substack{\text{order} \\ \text{cores}}}  
\end{split}  
\end{equation}  

In case of a plane wave incidence the right hand side vector $\bm{a}^{\text{inc}}$ of the Eq.~(\ref{eq:linear-system}) has a simple closed-form TT representation. The only non-zero elements are:
\begin{equation}
    \bm{a}^{\text{inc},-}_{0}(z_p) = \exp(i k_{z0} \Delta z_p)
\end{equation}
where $\Delta z_p = \left( N_S - p - 1/2 \right) h - H/2$. This vector can be treated as a Kronecker product:
\begin{equation}
    \begin{bmatrix}
        0 \\ 1
    \end{bmatrix}
    \otimes
    \begin{bmatrix}
        e^{i k_{z0} (N_S - h/2)} \\
        e^{i k_{z0} (N_S - 3h/2)} \\
        \vdots \\
        e^{i k_{z0} h/2}
    \end{bmatrix}
    \otimes
    \begin{bmatrix}
        0 \\ \vdots \\ 0 \\ 1 \\ 0 \\\vdots \\ 0
    \end{bmatrix}
\end{equation}
where the vector in the middle is of form $\exp(\alpha n + \varphi)$ and it's TT form can be found following the discussion of the section~\ref{ssec:elementary}. The last vector has $1$ in the position which corresponds to Fourier harmonic with $n=0$. Its QTT representation can be found in a similar way to the identity matrix. It also has unit ranks ($r_k = 1$), and its cores depend on the binary representation of index that represents the $n=0$ Fourier harmonic: $\mathcal{E}_k = [1, 0]$ if the $k$-th digit of this index in binary is 0, and $\mathcal{E}_k = [0, 1]$ otherwise. This construction ensures that the tensor train contraction evaluates to 1 only for the position $n=0$ (where all binary indices align) and 0 elsewhere.

\subsection{Amplitude conversion}
Conversion between plane wave amplitudes and Fourier amplitudes of the electric field projection $E_{ym}$ is defined by the matrices $\bm{X}$ and $\bm{Y}$. Given the identity matrix in the TT format, the matrix $\bm{X}$ can be constructed as a Kronecker product:
\begin{equation}
    \bm{X} = 
    \begin{bmatrix}
        1 & 1 \\
        1 & 1
    \end{bmatrix} 
    \otimes \bm{I}
\end{equation}

Matrix $\bm{Y}$ no longer has a closed form TT representation. However, it has only $2^{d_F}$ unique elements: one for each Fourier harmonic. A vector of this size is manageable enough to allow for its full calculation. The TT representation of such a vector can subsequently be obtained using the TT-SVD method. This approach guarantees that the predefined accuracy is preserved, as TT-SVD provides an analytical upper bound for the error.

The matrix $\bm{Y}$ is diagonal with non-zero elements:
\begin{equation}
    \bm{Y}_{pn, pn} = \frac{1}{k_{zn}}
\end{equation}
The $2^{d_F}$ of its unique elements are repeated on the diagonal. When these elements are calculated in TT form, the obtained vector should be tiled as shown in the section~\ref{ssec:kron} and turned into a diagonal matrix as shown in~\ref{ssec:diag}.

\subsection{Emission}
The matrix $\bm{D}$ describing the generalized source emission in each slice can be constructed in a similar way to the matrix $\bm{Y}$. It is block-diagonal with $2 N_S$ equal Toeplitz blocks:
\begin{equation}
    \bm{D}^\text{block}_{nm} = \varepsilon_{n-m}
\end{equation}
All considered Fourier harmonics of the dielectric permittivity $\varepsilon_n$ can be computed analytically and turned into a TT vector by the TT-SVD. A Toeplitz block $\bm{D}^\text{block}$ then can be obtained in the TT format in accordance with Section~\ref{ssec:diag}. This matrix then should be tiled $2 N_S$ times and turned into a block-diagonal matrix $\bm{D}$ as described in Sections \ref{ssec:kron} and~\ref{ssec:diag}.

\subsection{Wave propagation within a grating layer}
The remaining matrices $\bm{P}_b$ and $\bm{T}$, which describe redistribution of plane waves between grating slices, have more complicated structure and were treated numerically using the TT-Cross. As mentioned before, individual structural features had to be exploited as TT-Cross is only applicable to smooth tensors.

The matrix $\bm{P}_b$ represents propagation of the waves emitted by the generalized currents in each slice. It consists of two blocks:
\begin{equation}
    \bm{P}_b =
    \begin{bmatrix}
        \bm{P}_b^{+} & \textbf{0} \\
        \textbf{0} & \bm{P}_b^{-}
    \end{bmatrix}
\end{equation}
where:
\begin{equation}
    \left( \bm{P}_b^{+} \right)_{\! nm}(p, q) = 
    \delta_{nm} \cdot \begin{cases}
        e^{i k_{zn} h (p - q)}, & p > q \\
        \frac{1}{2} , & p = q \\
        0, & p < q
    \end{cases}
\label{equation:r-plus}
\end{equation}
and $\bm{P}_b^{-} = \left( \bm{P}_b^{+} \right)^T$. Ones $\bm{P}_b^{+}$ is calculated, the full matrix:
\begin{equation}
    \bm{P}_b =
    \begin{bmatrix}
        1 & 0 \\
        0 & 0
    \end{bmatrix}
    \otimes \bm{P}_b^{+}
    +
    \begin{bmatrix}
        0 & 0 \\
        0 & 1
    \end{bmatrix}
    \otimes \left( \bm{P}_b^{+} \right)^T
\end{equation}

To simplify computation of $\bm{P}_b^{+}$ the indices $n$ and $p$ can be permuted, i.e. the order indices $n_k$ will change slower than $p_k$. This permutation turns $\bm{P}_b^{+}$ into a block-diagonal matrix with Toeplitz blocks $( \bm{P}_b^{+} )_n$ being:
\begingroup
\renewcommand*{\arraystretch}{2.6}
\begin{equation}
\begin{bmatrix}
    \frac{1}{2}&0&\ldots&\ldots&0 \\
    e^{ik_{zn}h}&\ddots&\ddots&&\vdots \\
    e^{2ik_{zn}h}&\ddots&\ddots&\ddots&\vdots \\
    \vdots&\ddots&\ddots&\ddots& 0\\
    e^{(N_S-1)ik_{zn}h}&\ldots&e^{2ik_{zn}h}&e^{ik_{zn}h}&~~\frac{1}{2}~~
\end{bmatrix}
\label{equation:r-block}
\end{equation}
\endgroup
The $\bm{P}_b^{+}$ in this form can be constructed in four steps. First, we compute the tensor of exponent matrix blocks as a Kronecker product:
\begin{equation}
ih \cdot
\begin{bmatrix}
    0&0&\ldots&\ldots&0 \\
    1&\ddots&\ddots&&\vdots \\
    2&\ddots&\ddots&\ddots&\vdots \\
    \vdots&\ddots&\ddots&\ddots& 0\\
    N_S-1&\ldots&2&1&0
\end{bmatrix}
\otimes
\begin{bmatrix}
    k_{z,-\lfloor N_F/2 \rfloor} \\
    \vdots \\
    k_{z,\lfloor N_F/2 \rfloor}
\end{bmatrix}
\end{equation}
where both the matrix and vector are in the TT format. The matrix can be analytically decomposed into TT format as a lower triangular Toeplitz matrix, and the vector $\bm{k}_{z}$ can be defined using the TT-SVD as previously.

Second, the exponentiation function should be applied element-wise by the TT-Cross. Below the main diagonal the resulting tensor is the same as the one given by Eq.~(\ref{equation:r-block}). But on the main diagonal and higher there are ones instead of $1/2$ and zeros, respectively. So, as a third step, we subtract the Toeplitz matrix defined by anti-diagonal vector $\left[0,\, \ldots,\, 0,\, 1/2,\, 1,\, \ldots,\, 1\right]$ from each block. Finally, the computed tensor of diagonal blocks of $\bm{P}_b^{+}$ has to be turned into a block-diagonal matrix as shown in~(\ref{equation:block-diag-cores}).

Since $\bm{P}_b^{+}$ is constructed with permuted indices $p_k$ and $n_k$, we need to swap them back. The TT framework provides an index permutation algorithm, but it is relatively slow. The fastest way to do the permutation is to swap order and slice cores in the first step, and then perform all operations with permuted indices. For example, the block diagonal matrix construction algorithm should be modified so that zero layers are added to the last cores instead of the first ones.

The matrix $\bm{T}$ is a non-square matrix of size $2N_F \times 2N_F N_S$. When multiplied by the amplitude vector of the self-consistent field calculated in each slice, it gives a vector of the resulting outgoing field amplitudes at the grating layer boundaries $z=\pm H/2$:
\begin{equation}
    \bm{T} =
    \begin{bmatrix}
        \textbf{0} & \bm{T}^{+} \\
        \bm{T}^{-} & \textbf{0}
    \end{bmatrix}
\end{equation}
The blocks $\bm{T}^{\pm}$ are in anti-diagonal positions.
\begin{equation}
    \bm{T}^{\pm}_{nm}(p) = \delta_{nm}\exp(ik_{zn}\Delta^{\pm}_p)
\end{equation}
where $\Delta^{+}_p = h (N_S - p + 1/2)$, $\Delta^{-}_p = h (p - 1/2)$.
The construction of this matrix in the TT format is almost identical to $\bm{P}_b$ except that $\bm{T}$ is non-square, so the tensor mode sizes corresponding to the slice indices are set to one.

It's important to note that there is no guarantee that the TT-Cross algorithm used to perform exponentiation will work stably with arbitrarily large $d_F$ and $d_S$. While there is no formal criterion for the level of smoothness required for accurate TT-Cross evaluations, the algorithm generally performs well in practice. The instability is unlikely to arise in most practical applications. In fact, the issue only becomes significant when an excessive number of evanescent Fourier harmonics are considered -- something that is not necessary in practical scenarios.

In such cases, some of the matrices $(\bm{P}_b^{+})_{n}$ may become predominantly zero due to the decay of the excessive evanescent waves within a single grating slice. If this issue does occur, a slightly more robust implementation of the method involves computing a smoother symmetrized version of the matrix, $\bm{P}_b^{+} + (\bm{P}_b^{+})^T$, and extracting its lower-diagonal part. The same should be done to the $\bm{T}$ matrix.

\section{Numerical examples}
To check that the accelerated method works correctly, its results were compared with the conventional FMM and unmodified GSM. The results agreed within a given accuracy. Fig.~\ref{fig:accuracy-control-plot} demonstrates the convergence of the new GSM implementation to benchmark results with three different preset tolerances, namely $10^{-3}$, $10^{-6}$, and $10^{-9}$. This tolerance is an accuracy of the TT-rounding algorithm and a tolerance of the linear system solver AMEn, which both were set to be the same. The benchmark solution for a wavelength-scale dielectric lamellar grating was obtained by the conventional FMM by verifying that it converged to an accuracy better than $10^{-10}$. One can see, that the new TT-accelerated implementation reveals even better accuracy than the preset tolerance.

\begin{figure}[!htb]
    \centering
    \small
    \begin{tikzpicture}[scale = 1.0]
        \definecolor{clr1}{RGB}{0,114,178}
        \definecolor{clr2}{RGB}{240,105,0}
        \definecolor{clr3}{RGB}{0,158,115}
        
        \begin{axis}[
            width = \linewidth,
            height = 0.7\linewidth,
            xlabel = {Number of harmonics},
            ylabel = {Difference with FMM},
            minor tick num = 3,
            xmin = 1,
            xmax = 1e5,
            xmode=log,
            ymin = 2e-11,
            ymax = 1,
            ymode=log,
            grid = major,
            legend pos = south west
        ]   
            \addplot[
                color=clr3, 
                mark=*,
                mark options = {scale = 0.6},
                smooth,
                line width = 0.4mm
            ]
            table[x index=0, y index=1]
            {
                4	8.95E-02
                8	1.57E-02
                16	3.46E-03
                32	8.27E-04
                64	2.03E-04
                128	5.04E-05
                256	1.26E-05
                512	3.13E-06
                1024	7.83E-07
                2048	1.96E-07
                4096	4.92E-08
                8192	1.24E-08
                16384	2.91E-09
                32768	1.75E-10
                65536	4.80E-10
            };
            \addlegendentry{$10^{-9}$}
        
            \addplot[
                color=clr2, 
                mark=*,
                mark options = {scale = 0.6},
                smooth,
                line width = 0.4mm
            ]
            table[x index=0, y index=1]
            {
                4	8.95E-02
                8	1.57E-02
                16	3.46E-03
                32	8.27E-04
                64	2.03E-04
                128	5.05E-05
                256	1.25E-05
                512	3.15E-06
                1024	7.11E-07
                2048	2.42E-07
                4096	7.39E-08
                8192	2.17E-09
                16384	1.30E-07
                32768	1.51E-08
                65536	3.35E-08
            };
            \addlegendentry{$10^{-6}$}
    
            \addplot[
                color=clr1, 
                mark=*,
                mark options = {scale = 0.6},
                smooth,
                line width = 0.4mm
            ]
            table[x index=0, y index=1]
            {
                4	8.95E-02
                8	1.57E-02
                16	3.80E-03
                32	8.47E-04
                64	2.90E-04
                128	1.98E-04
                256	1.03E-04
                512	9.14E-05
                1024	4.24E-05
                2048	9.28E-06
                4096	4.10E-05
                8192	2.95E-05
                16384	7.15E-05
                32768	1.39E-05
                65536	7.25E-05
            };
            \addlegendentry{$10^{-3}$}
            
            \addplot[clr3, domain = 1:10^6, dashed] {1e-9};
            \addplot[clr2, domain = 1:10^6, dashed] {1e-6};
            \addplot[clr1, domain = 1:10^6, dashed] {1e-3};
        \end{axis}
    \end{tikzpicture}
    \caption{Convergence of the new algorithm (GSM-TT) with respect to the number of  Fourier harmonics $N_F$ considered to a benchmark solution obtained by the FMM. Three different preset tolerances are shown, namely $10^{-3}$, $10^{-6}$, and $10^{-9}$, and are represented with dashed lines. The benchmark structure is a wavelength-scale dielectric lamellar grating.}
    \label{fig:accuracy-control-plot}
\end{figure}

Then, given a fixed preset tolerance, we have evaluated computational performance of the new GSM implementation. Fig.~\ref{fig:runtime-memory-plot} compares the algorithm runtime and memory usage with unmodified GSM. Number of slices $N_S$ has been set to be equal to the number of Fourier harmonics $N_F$.

\begin{figure}[!htb]
    \centering
    \small
    \begin{tikzpicture}[scale = 1.0]
        \definecolor{clr1}{RGB}{0,114,178}
        \definecolor{clr2}{RGB}{240,105,0}
        \definecolor{clr3}{RGB}{0,158,115}
        
        \begin{axis}[
            width = \linewidth,
            height = 0.6\linewidth,
            xlabel = {Number of harmonics},
            ylabel = {Runtime, s},
            minor tick num = 3,
            xmin = 4,
            xmax = 100000,
            xmode=log,
            ymin = 0,
            ymax = 150,
            grid = major,
            legend pos = north west
        ]
            \addplot[
                color=clr2, 
                mark=*,
                mark options = {scale = 0.6},
                line width = 0.4mm,
                smooth
            ]
            table[x index=0, y index=1]
            {
                4	0.2408228
                8	0.3452857
                16	0.5955368
                32	1.2627230
                64	1.5719710
                128	2.6264346
                256	4.1573390
                512	8.7080634
                1024	19.0249052
                2048	32.6166938
                4096	52.9303816
                8192	69.1035078
                16384	90.6937472
                32768	110.7450733
                65536	133.5077095
            };
            \addlegendentry{GSM-TT}
    
            \addplot[clr1, domain = 1:2^13, smooth, line width = 0.4mm] {0.02925989*x};
            \addlegendentry{GSM}
        \end{axis}
    \end{tikzpicture}
    
    \bigskip
    
    \begin{tikzpicture}[scale = 1.0]
        \definecolor{clr1}{RGB}{0,114,178}
        \definecolor{clr2}{RGB}{240,105,0}
        \definecolor{clr3}{RGB}{0,158,115}
        
        \begin{axis}[
            width = \linewidth,
            height = 0.6\linewidth,
            xlabel = {Number of harmonics},
            ylabel = {Memory usage, GB},
            minor tick num = 3,
            xmin = 5,
            xmax = 100000,
            xmode=log,
            ymin = 0,
            ymax = 5,
            grid = major,
            legend pos = north west
        ]
            \addplot[
                color=clr2, 
                mark=*,
                mark options = {scale = 0.6},
                smooth,
                line width = 0.4mm
            ]
            table[x index=0, y index=1]
            {
                8	0.0238
                16	0.0426
                32	0.0798
                64	0.1225
                128	0.2394
                256	0.4028
                512	0.6006
                1024	0.9444
                2048	1.3938
                4096	1.7760
                8192	2.0913
                16384	2.7476
                32768	3.1607
                65536	3.8979
            };
            \addlegendentry{GSM-TT}
            
            \addplot[clr1, domain = 1:2^12, smooth, line width = 0.4mm] {0.004960914*x};
            \addlegendentry{GSM}
            
        \end{axis}
    \end{tikzpicture}
    \caption{Running time (above) and memory usage (below) of the new GSM-TT implementation compared to the unmodified GSM.}
    \label{fig:runtime-memory-plot}
\end{figure}

The results shown should only be considered from an asymptotic point of view, as the methods are implemented differently. However, it is clear that the GSM-TT requires exponentially less time and memory than conventional methods. This allows a much larger number of spatial field harmonics to be considered, and thus diffraction by more complex structures to be simulated.

To test capabilities of the method for simulation of multi-scale structures of different thicknesses, we considered binary gratings consisting of multiple pixels with different random widths (as illustrated in Fig.~\ref{fig:grating}). Distance between the centers of all pixels is fixed and equal to the incident wavelength $\lambda$. Fig.~\ref{fig:multiscale-convergence} shows the convergence of the new implementation with gratings periods ranging from $\lambda$ to $1000\lambda$. Fig.~\ref{fig:thickness-sweep} demonstrates that the convergence for gratings with a fixed period (here $10\lambda$) is practically the same for various grating thickness. When enough harmonics are considered, all simulations converge to an accuracy not exceeding the TT rounding value equal to $10^{-6}$. The rate of convergence resembles the slope $O(x^{-2})$.

\begin{figure}[!htb]
    \centering
    \small
    
    \pgfplotsset{
        colormap={bluegradient}{hsb=(0.56,0.5,0.9) hsb=(0.56,1,0.7) hsb=(0.56,1,0.5) hsb=(0.56,1,0)}
    }
    \definecolor{myorange}{RGB}{240,105,0}
    
    \begin{tikzpicture}[scale = 1.0] 
        \begin{axis}[
            width = \linewidth,
            height = 0.8\linewidth,
            ylabel near ticks,
            xlabel near ticks,
            xlabel = {Number of harmonics},
            ylabel = {Difference between steps},
            minor tick num = 3,
            xmin = 5,
            xmax = 50000,
            xmode=log,
            ymax = 1,
            ymin = 1e-11,
            ymode=log,
            grid = major,
            legend pos = south west,
            colormap name=bluegradient,
            cycle list={[of colormap]},
            every axis plot/.append style={
                mark=*, line width=1pt, mark options={scale = 0.6}
            }
        ]
            \addplot[mark=none, line width=0.5pt, myorange, domain = 1:10^5, smooth, dashed, forget plot] {1/x^2};
            
            \addplot table[x index=0, y index=1]
            {
                8       1.6135E-04
                16      3.5177E-06
                32      3.9906E-07
                64      5.1082E-07
                128	    2.0143E-07
                256	    5.8322E-08
                512	    9.5400E-09
                1024	8.4696E-09
                2048	1.4823E-09
                4096	7.3456E-10
                8192	1.2368E-10
                16384	6.5978E-10
                32768	5.7683E-10
            };
            \addlegendentry{$1 \lambda$}
    
            \addplot table[x index=0, y index=1]
            {
                8	5.5753E-04
                16	2.7577E-04
                32	1.0085E-02
                64	3.8152E-05
                128	4.2738E-06
                256	1.7275E-06
                512	2.6293E-07
                1024	4.3742E-08
                2048	1.8423E-08
                4096	7.7124E-09
                8192	1.4892E-08
                16384	1.0476E-08
                32768	1.7234E-08
            };
            \addlegendentry{$10 \lambda$}
    
            \addplot table[x index=0, y index=1]
            {
                8	2.6332E-05
                16	4.9582E-05
                32	1.2961E-04
                64	5.8605E-04
                128	3.7209E-04
                256	1.0788E-02
                512	8.8640E-06
                1024	2.7100E-05
                2048	4.5899E-06
                4096	6.5316E-07
                8192	5.9245E-08
                16384	2.6537E-08
                32768	2.3944E-08
            };
            \addlegendentry{$100 \lambda$}
    
            \addplot table[x index=0, y index=1]
            {
                8	3.5509E-05
                16	4.6539E-06
                32	2.0681E-05
                64	7.1683E-05
                128	7.7536E-05
                256	1.8402E-04
                512	3.1516E-04
                1024	5.2368E-04
                2048	1.0738E-02
                4096	8.3247E-05
                8192	7.9290E-05
                16384	7.4724E-06
                32768	7.0807E-08
            };
            \addlegendentry{$1000 \lambda$}
        \end{axis}
    \end{tikzpicture}
    \caption{Convergence (absolute difference between zero harmonic amplitudes for subsequent $N_F$) of the new GSM implementation for two-scale gratings. The grating sub-pixel size equals to the wavelength $\lambda$, and the macro-period varies from $\lambda$ to $1000\lambda$. Grating thickness is set to $0.5\lambda$. Dashed line indicates the functional dependence $O(x^{-2})$.}
    \label{fig:multiscale-convergence}
\end{figure}

\begin{figure}[!htb]
    \centering
    \small

    \pgfplotsset{
        colormap={bluegradient}{hsb=(0.56,0.3,0.95) hsb=(0.56,1,0.7) hsb=(0.56,1,0)}
    }
    \definecolor{myorange}{RGB}{240,105,0}
    \def\thpt{0.7pt}
    
    \begin{tikzpicture}[scale = 1.0] 
        \begin{axis}[
            width = \linewidth,
            height = 0.7\linewidth,
            ylabel near ticks,
            xlabel near ticks,
            xlabel = {Number of harmonics},
            ylabel = {Difference between steps},
            minor tick num = 3,
            xmin = 5,
            xmax = 50000,
            xmode=log,
            ymax = 1,
            ymin = 1e-10,
            ymode=log,
            grid = major,
            point meta min=0,
            point meta max=1,
            colorbar horizontal,
            colorbar sampled,
            colorbar style={
                samples=11,
                xlabel={Thickness relative to wavelength},
                height=6pt,
                xtick={.05, .15, .25, .35, .45, .55, .65, .75, .85, .95},
                xticklabels={0.1, 0.2, 0.3, 0.4, 0.5, 0.5, 0.6, 0.8, 0.9, 1.0},
                tick style={draw=none},
                tick label style={font=\scriptsize}},
        ]
            \addplot[mark=none, line width=0.5pt, myorange, domain = 1:10^5, smooth, dashed] {1/x^2};
            
            \addplot[mesh, point meta=0.1, line width=1pt] table[x index=0, y index=1]
            {
                8	9.125E-04
                16	3.766E-04
                32	1.799E-02
                64	4.975E-05
                128	4.032E-07
                256	2.372E-06
                512	3.780E-07
                1024	5.299E-09
                2048	3.888E-08
                4096	1.741E-08
                8192	1.342E-08
                16384	9.803E-09
                32768	1.813E-08
            };
    
            \addplot[mesh, point meta=0.2, line width=\thpt] table[x index=0, y index=1]
            {
                8	1.118E-03
                16	1.682E-04
                32	5.768E-02
                64	2.710E-03
                128	1.495E-04
                256	1.569E-05
                512	1.900E-06
                1024	2.826E-07
                2048	2.093E-08
                4096	9.029E-09
                8192	1.194E-08
                16384	4.670E-08
                32768	5.044E-08
            };
    
            \addplot[mesh, point meta=0.3, line width=\thpt] table[x index=0, y index=1]
            {
                8	2.356E-03
                16	7.806E-04
                32	6.812E-02
                64	4.047E-03
                128	1.399E-04
                256	1.069E-05
                512	6.048E-07
                1024	1.048E-07
                2048	1.601E-08
                4096	3.844E-08
                8192	1.242E-07
                16384	9.609E-09
                32768	1.093E-08
            };
    
            \addplot[mesh, point meta=0.4, line width=\thpt] table[x index=0, y index=1]
            {
                8	5.885E-03
                16	2.854E-03
                32	1.025E-01
                64	1.467E-03
                128	1.406E-04
                256	2.099E-05
                512	2.902E-06
                1024	6.426E-07
                2048	1.797E-07
                4096	2.709E-07
                8192	1.268E-07
                16384	6.926E-08
                32768	2.605E-08
            };
    
            \addplot[mesh, point meta=0.5, line width=\thpt] table[x index=0, y index=1]
            {
                8	7.701E-03
                16	3.247E-03
                32	1.480E-01
                64	1.899E-03
                128	2.693E-04
                256	3.832E-05
                512	6.667E-06
                1024	1.585E-06
                2048	5.089E-07
                4096	4.986E-07
                8192	4.168E-07
                16384	2.721E-07
                32768	2.482E-07
            };
    
            \addplot[mesh, point meta=0.6, line width=\thpt] table[x index=0, y index=1]
            {
                8	4.125E-03
                16	1.074E-03
                32	1.678E-01
                64	5.347E-03
                128	4.748E-04
                256	6.292E-05
                512	1.058E-05
                1024	1.947E-06
                2048	4.667E-07
                4096	1.472E-07
                8192	1.062E-07
                16384	4.847E-08
                32768	7.807E-08
            };
    
            \addplot[mesh, point meta=0.7, line width=\thpt] table[x index=0, y index=1]
            {
                8	7.400E-03
                16	3.382E-03
                32	1.653E-01
                64	1.000E-02
                128	4.777E-04
                256	7.070E-05
                512	1.150E-05
                1024	2.578E-06
                2048	8.575E-08
                4096	3.118E-07
                8192	1.642E-07
                16384	5.505E-08
                32768	4.000E-08
            };
    
            \addplot[mesh, point meta=0.8, line width=\thpt] table[x index=0, y index=1]
            {
                8	1.775E-02
                16	9.178E-03
                32	1.838E-01
                64	1.698E-02
                128	6.151E-04
                256	1.623E-05
                512	1.261E-05
                1024	3.655E-06
                2048	1.745E-06
                4096	2.310E-07
                8192	6.893E-10
                16384	1.876E-07
                32768	6.378E-08
                
            };
    
            \addplot[mesh, point meta=0.9, line width=\thpt] table[x index=0, y index=1]
            {
                8	3.480E-03
                16	5.398E-03
                32	2.145E-01
                64	6.867E-03
                128	1.490E-03
                256	5.357E-05
                512	2.125E-05
                1024	6.101E-06
                2048	7.244E-07
                4096	1.145E-06
                8192	1.093E-07
                16384	1.084E-07
                32768	2.706E-07
            };
    
            \addplot[mesh, point meta=1, line width=\thpt] table[x index=0, y index=1]
            {
                8	2.191E-02
                16	7.373E-04
                32	2.167E-01
                64	9.295E-03
                128	5.511E-04
                256	8.228E-05
                512	2.388E-05
                1024	6.787E-06
                2048	9.908E-07
                4096	9.623E-07
                8192	6.616E-08
                16384	5.454E-08
                32768	1.710E-07
            };
        \end{axis}
    \end{tikzpicture}
    \caption{Convergence of the new method for two-scale gratings with period $10\lambda$. Grating thickness varies in range from $0.1\lambda$ to $\lambda$. The value on vertical axis is the difference between zero harmonic amplitudes calculated with adjacent $d_F$. The dashed line indicates the functional dependence $O(x^{-2})$.}
    \label{fig:thickness-sweep}
\end{figure}

An important characteristic of the developed method is the ranks of the TT decomposition  of all matrices in the linear system. Analysis of these ranks provides a valuable insight into where compression is achieved most effectively.

Matrices $\bm{I}$, $\bm{X}$ and vector $\vb{a}^\text{inc}$ with closed-form TT representations require very small TT ranks, not exceeding $r = 4$, which remain independent of the grating or incidence parameters. Their size in the TT format is negligible compared to other matrices in the system.

The matrices $\bm{Y}$, $\bm{P}_b$, and $\bm{T}$ require significantly larger ranks that depend on the grating period and the number of Fourier harmonics $N_F$. For $\bm{P}_b$ and $\bm{T}$, the ranks also depend on the grating thickness and the number of slices, $N_S$. Although the exact dependency is complex, in all the examples presented in this paper, the ranks do not exceed a few dozen. Therefore, these matrices remain manageable in the TT format.

The Toeplitz matrix $\bm{D}$ exhibits the highest TT ranks among all the matrices. These ranks remain independent of the incidence and grating thickness. Fig.~\ref{fig:plot-pb-ranks} illustrates how the ranks vary with the number of considered Fourier harmonics $N_F$ and the number of wavelength-scale pixels within the grating macro-period. As can be seen from the plot, the ranks of $\bm{D}$ increase with $N_F$, but eventually plateau. The maximum rank is linearly dependent on the number of pixels in the grating. This suggests that the ranks are proportional to the number of degrees of freedom in the grating shape.

\begin{figure}[!htb]
    \centering
    \small
    
    \pgfplotsset{
        colormap={bluegradient}{hsb=(0.56,0.3,0.95) hsb=(0.56,1,0.7) hsb=(0.56,1,0)}
    }
    \def\thpt{1pt}
    
    \begin{tikzpicture}[scale = 1.0] 
        \begin{axis}[
            width = \linewidth,
            height = 0.7\linewidth,
            ylabel near ticks,
            xlabel near ticks,
            xlabel = {Number of harmonics},
            ylabel = {TT rank of $\vb*{D}$},
            minor tick num = 3,
            xmin = 100,
            xmax = 1048576,
            xmode=log,
            ymax = 500,
            ymin = 1,
            grid = major,
            point meta min=0,
            point meta max=110,
            colorbar horizontal,
            colorbar sampled,
            colorbar style={
                samples=12,
                xlabel={Period relative to wavelength},
                height=6pt,
                xtick={5, 15, 25, 35, 45, 55, 65, 75, 85, 95, 105},
                xticklabels={1, 10, 20, 30, 40, 50, 60, 70, 80, 90, 100},
                tick style={draw=none},
                tick label style={font=\scriptsize}},
        ]        
            \addplot[mesh, point meta=100, line width=\thpt] table[x index=0, y index=1]{
                4       3
                8       3
                16      7
                32      7
                64      15
                128     15
                256     31
                512     31
                1024    63
                2048    63
                4096    127
                8192    127
                16384   250
                32768   255
                65536   373
                131072  416
                262144  431
                524288  439
                1048576 446
            };
            \addplot[mesh, point meta=90, line width=\thpt] table[x index=0, y index=1]{
                4       3
                8       3
                16      7
                32      7
                64      15
                128     15
                256     31
                512     31
                1024    63
                2048    63
                4096    127
                8192    127
                16384   244
                32768   255
                65536   340
                131072  380
                262144  394
                524288  404
                1048576 412
            };
            \addplot[mesh, point meta=80, line width=\thpt] table[x index=0, y index=1]{
                4       3
                8       3
                16      7
                32      7
                64      15
                128     15
                256     31
                512     31
                1024    60
                2048    63
                4096    127
                8192    127
                16384   228
                32768   255
                65536   323
                131072  352
                262144  363
                524288  369
                1048576 372
            };
            \addplot[mesh, point meta=70, line width=\thpt] table[x index=0, y index=1]{
                4       3
                8       3
                16      7
                32      7
                64      15
                128     15
                256     31
                512     31
                1024    63
                2048    63
                4096    127
                8192    127
                16384   232
                32768   254
                65536   292
                131072  311
                262144  325
                524288  332
                1048576 336
            };
            \addplot[mesh, point meta=60, line width=\thpt] table[x index=0, y index=1]{
                4       3
                8       3
                16      7
                32      7
                64      15
                128     15
                256     31
                512     31
                1024    63
                2048    63
                4096    127
                8192    127
                16384   219
                32768   247
                65536   262
                131072  280
                262144  291
                524288  295
                1048576 298
            };
            \addplot[mesh, point meta=50, line width=\thpt] table[x index=0, y index=1]{
                4       3
                8       3
                16      7
                32      7
                64      15
                128     15
                256     31
                512     31
                1024    63
                2048    63
                4096    124
                8192    127
                16384   202
                32768   227
                65536   236
                131072  245
                262144  250
                524288  252
                1048576 255
            };
            \addplot[mesh, point meta=40, line width=\thpt] table[x index=0, y index=1]{
                4       3
                8       3
                16      7
                32      7
                64      15
                128     15
                256     31
                512     31
                1024    63
                2048    63
                4096    120
                8192    127
                16384   180
                32768   198
                65536   203
                131072  206
                262144  207
                524288  210
                1048576 212
            };
            \addplot[mesh, point meta=30, line width=\thpt] table[x index=0, y index=1]{
                4       3
                8       3
                16      7
                32      7
                64      15
                128     15
                256     31
                512     31
                1024    63
                2048    63
                4096    119
                8192    126
                16384   144
                32768   158
                65536   163
                131072  165
                262144  165
                524288  166
                1048576 167
            };
            \addplot[mesh, point meta=20, line width=\thpt] table[x index=0, y index=1]{
                4       3
                8       3
                16      7
                32      7
                64      15
                128     15
                256     31
                512     31
                1024    59
                2048    63
                4096    92
                8192    106
                16384   111
                32768   115
                65536   117
                131072  118
                262144  119
                524288  120
                1048576 120
            };
            \addplot[mesh, point meta=10, line width=\thpt] table[x index=0, y index=1]{
                4       3
                8       3
                16      7
                32      7
                64      15
                128     15
                256     31
                512     31
                1024    50
                2048    60
                4096    63
                8192    68
                16384   69
                32768   70
                65536   71
                131072  71
                262144  71
                524288  71
                1048576 71
            };
            \addplot[mesh, point meta=1, line width=\thpt] table[x index=0, y index=1]{
                4       3
                8       3
                16      7
                32      7
                64      12
                128     12
                256     12
                512     13
                1024    14
                2048    14
                4096    14
                8192    14
                16384   14
                32768   14
                65536   14
                131072  14
                262144  14
                524288  14
                1048576 14
            };
        \end{axis}
    \end{tikzpicture}
    \caption{Ranks of the Toeplitz matrix $\bm{D}$ as a function of the number of Fourier harmonics considered and the complexity of the grating geometry. The size of sub-pixels equals to the wavelength $\lambda$, and the grating macro-period ranges from $1\lambda$ to $100\lambda$.}
    \label{fig:plot-pb-ranks}
\end{figure}

\section{Conclusion}
To conclude, this work has accelerated the Generalized Source Method (GSM) by integrating adapted Tensor Train (TT) decomposition algorithms. The resulting approach efficiently solves large-period multiscale grating diffraction problems with $O(\log N)$ asymptotic complexity for both numerical computation and memory usage. Our code, which we developed for this research, can be found on the GitHub \cite{ttgsm}.

The results illustrate how tensor compression methods, such as TT decomposition, can enhance existing wavelength-scale electrodynamical computation techniques. While extending this approach to 2D metasurfaces or other computational methods in the field appears promising, challenges like TT-Cross instability require careful handling. Nonetheless, by addressing these challenges, the method can substantially reduce parameter redundancy in scattering problems, offering a valuable tool for specific electrodynamical applications.

\section*{Acknowledgements}
The work was supported by the Russian Science Foundation, grant No.22-11-00153. The authors are grateful to the Professor Ivan Oseledets for advice on the use of his codes.

\bibliographystyle{ieeetr}
\bibliography{tt-gsm}

\begin{thebibliography}{10}

\bibitem{antonakakis:hal-00785737}
T.~Antonakakis, F.~I. Baida, A.~Belkhir, K.~Cherednichenko, S.~Cooper,
  R.~Craster, G.~Dem{\'e}sy, J.~Desanto, G.~Granet, B.~Gralak, S.~Guenneau,
  D.~Maystre, A.~Nicolet, B.~Stout, F.~Zolla, B.~Vial, and E.~Popov, {\em
  {Gratings: Theory and Numeric Applications}}.
\newblock {AMU (PUP)}, Dec. 2012.

\bibitem{Schulz2024}
S.~A. Schulz, R.~F. Oulton, M.~Kenney, A.~Alù, I.~Staude, A.~Bashiri,
  Z.~Fedorova, R.~Kolkowski, A.~F. Koenderink, X.~Xiao, J.~Yang, W.~J. Peveler,
  A.~W. Clark, G.~Perrakis, A.~C. Tasolamprou, M.~Kafesaki, A.~Zaleska,
  W.~Dickson, D.~Richards, A.~Zayats, H.~Ren, Y.~Kivshar, S.~Maier, X.~Chen,
  M.~A. Ansari, Y.~Gan, A.~Alexeev, T.~F. Krauss, A.~Di~Falco, S.~D. Gennaro,
  T.~Santiago-Cruz, I.~Brener, M.~V. Chekhova, R.-M. Ma, V.~V. Vogler-Neuling,
  H.~C. Weigand, U.-L. Talts, I.~Occhiodori, R.~Grange, M.~Rahmani, L.~Xu,
  S.~M. Kamali, E.~Arababi, A.~Faraon, A.~C. Harwood, S.~Vezzoli, R.~Sapienza,
  P.~Lalanne, A.~Dmitriev, C.~Rockstuhl, A.~Sprafke, K.~Vynck, J.~Upham, M.~Z.
  Alam, I.~De~Leon, R.~W. Boyd, W.~J. Padilla, J.~M. Malof, A.~Jana, Z.~Yang,
  R.~Colom, Q.~Song, P.~Genevet, K.~Achouri, A.~B. Evlyukhin, U.~Lemmer, and
  I.~Fernandez-Corbaton, ``Roadmap on photonic metasurfaces,'' {\em Applied
  Physics Letters}, vol.~124, June 2024.

\bibitem{Bonnet2018}
M.~Bonnet, F.~Collino, E.~Demaldent, A.~Imperiale, and L.~Pesudo, ``A hybrid
  method combining the surface integral equation method and ray tracing for the
  numerical simulation of high frequency diffraction involved in ultrasonic
  ndt,'' {\em Journal of Physics: Conference Series}, vol.~1017, p.~012007, May
  2018.

\bibitem{Hudelist2009}
F.~Hudelist, A.~J. Waddie, and M.~R. Taghizadeh, ``Analysis of crossed gratings
  with large periods and small feature sizes by stitching of the
  electromagnetic field,'' {\em Journal of the Optical Society of America A},
  vol.~26, p.~2648, Nov. 2009.

\bibitem{Hughes2021}
T.~W. Hughes, M.~Minkov, V.~Liu, Z.~Yu, and S.~Fan, ``A perspective on the
  pathway toward full wave simulation of large area metalenses,'' {\em Applied
  Physics Letters}, vol.~119, Oct. 2021.

\bibitem{Salary2016}
M.~M. Salary, A.~Forouzmand, and H.~Mosallaei, ``Model order reduction of
  large-scale metasurfaces using a hierarchical dipole approximation,'' {\em
  ACS Photonics}, vol.~4, p.~63–75, Dec. 2016.

\bibitem{Gao2024}
H.-W. Gao, X.-M. Xin, Q.~J. Lim, S.~Wang, and Z.~Peng, ``Efficient full-wave
  simulation of large-scale metasurfaces and metamaterials,'' {\em IEEE
  Transactions on Antennas and Propagation}, vol.~72, p.~800–811, Jan. 2024.

\bibitem{10237050}
N.~Anselmi, G.~Oliveri, L.~Poli, A.~Polo, P.~Rocca, M.~Salucci, and A.~Massa,
  {\em Breaking the Curse of Dimensionality in Electromagnetics Design Through
  Optimization Empowered by Machine Learning}, pp.~81--104.
\newblock John Wiley \& Sons, Ltd, 2023.

\bibitem{Majorel2022}
C.~Majorel, C.~Girard, A.~Arbouet, O.~L. Muskens, and P.~R. Wiecha, ``Deep
  learning enabled strategies for modeling of complex aperiodic plasmonic
  metasurfaces of arbitrary size,'' {\em ACS Photonics}, vol.~9, p.~575–585,
  Jan. 2022.

\bibitem{An2020}
S.~An, B.~Zheng, M.~Y. Shalaginov, H.~Tang, H.~Li, L.~Zhou, J.~Ding, A.~M.
  Agarwal, C.~Rivero-Baleine, M.~Kang, K.~A. Richardson, T.~Gu, J.~Hu,
  C.~Fowler, and H.~Zhang, ``Deep learning modeling approach for metasurfaces
  with high degrees of freedom,'' {\em Optics Express}, vol.~28, p.~31932, Oct.
  2020.

\bibitem{Skarda2022}
J.~Skarda, R.~Trivedi, L.~Su, D.~Ahmad-Stein, H.~Kwon, S.~Han, S.~Fan, and
  J.~Vučković, ``Low-overhead distribution strategy for simulation and
  optimization of large-area metasurfaces,'' {\em npj Computational Materials},
  vol.~8, Apr. 2022.

\bibitem{10221390}
V.~Medvedev, A.~Erdmann, and A.~Rosskopf, ``Modeling of near-and far-field
  diffraction from euv absorbers using physics-informed neural networks,'' in
  {\em 2023 Photonics \& Electromagnetics Research Symposium (PIERS)},
  pp.~297--305, 2023.

\bibitem{Sarkar2023}
S.~Sarkar, A.~Ji, Z.~Jermain, R.~Lipton, M.~Brongersma, K.~Dayal, and H.~Y.
  Noh, ``Physics‐informed machine learning for inverse design of optical
  metamaterials,'' {\em Advanced Photonics Research}, vol.~4, Oct. 2023.

\bibitem{Petit1980}
R.~Petit, ed., {\em Electromagnetic Theory of Gratings}.
\newblock Springer Berlin Heidelberg, 1980.

\bibitem{SHCHERBAKOV-2012-2DGSM}
A.~A. Shcherbakov and A.~V. Tishchenko, ``New fast and memory-sparing method
  for rigorous electromagnetic analysis of 2d periodic dielectric structures,''
  {\em Journal of Quantitative Spectroscopy and Radiative Transfer}, vol.~113,
  no.~2, pp.~158--171, 2012.

\bibitem{Chandezon2024}
J.~Chandezon and G.~Granet, ``Diffraction by gratings: from the c-method to the
  stochastic c-method,'' {\em Journal of the Optical Society of America A},
  vol.~41, p.~1675, Aug. 2024.

\bibitem{SHCHERBAKOV-2010-1DGSM}
A.~Shcherbakov and A.~Tishchenko, ``Fast numerical method for modelling
  one-dimensional diffraction gratings,'' {\em Quantum Electronics}, vol.~40,
  p.~538, 08 2010.

\bibitem{Shcherbakov2017}
A.~Shcherbakov, O.~Shavdina, A.~Tishchenko, C.~Veillas, I.~Verrier, O.~Dellea,
  and Y.~Jourlin, ``Optical diffraction by ordered 2d arrays of silica
  microspheres,'' {\em Journal of Quantitative Spectroscopy and Radiative
  Transfer}, vol.~189, p.~37–42, Mar. 2017.

\bibitem{OSELEDETS-2011-TT}
I.~V. Oseledets, ``Tensor-train decomposition,'' {\em SIAM Journal on
  Scientific Computing}, vol.~33, no.~5, pp.~2295--2317, 2011.

\bibitem{DOLGOV-2014-AMEN}
S.~V. Dolgov and D.~V. Savostyanov, ``Alternating minimal energy methods for
  linear systems in higher dimensions,'' {\em {SIAM} Journal on Scientific
  Computing}, vol.~36, pp.~A2248--A2271, jan 2014.

\bibitem{TISHCHENKO-2000-GSM}
A.~V. Tishchenko, ``Generalized source method: New possibilities for waveguide
  and grating problems,'' {\em Optical and Quantum Electronics}, vol.~32,
  pp.~971--980, Aug 2000.

\bibitem{chew1995waves}
W.~Chew, {\em Waves and Fields in Inhomogeneous Media}.
\newblock Electromagnetic Waves Series, IEEE Press, 1995.

\bibitem{ttpy}
I.~Oseledets, ``{TTPy 1.0}.'' \url{https://github.com/oseledets/ttpy}, 2015.

\bibitem{Gantmacher-1959-Matrices}
H.~Schwerdtfeger, ``The theory of matrices, by f. r. gantmacher. chelsea, new
  york, 1959. vol. i. x 374 pages; vol. ii. ix 276 pages. \$6.00 per volume.,''
  {\em Canadian Mathematical Bulletin}, vol.~4, no.~1, p.~82–85, 1961.

\bibitem{Goreinov-1997-pseudo-skeleton}
S.~A. Goreinov, E.~E. Tyrtyshnikov, and N.~Zamarashkin, ``A theory of
  pseudoskeleton approximations,'' {\em Linear Algebra and its Applications},
  vol.~261, pp.~1--21, 1997.

\bibitem{OSELEDETS2010TTCROSS}
I.~Oseledets and E.~Tyrtyshnikov, ``Tt-cross approximation for multidimensional
  arrays,'' {\em Linear Algebra and its Applications}, vol.~432, no.~1,
  pp.~70--88, 2010.

\bibitem{Kazeev-2013-TTToeplitz}
V.~A. Kazeev, B.~N. Khoromskij, and E.~E. Tyrtyshnikov, ``Multilevel toeplitz
  matrices generated by tensor-structured vectors and convolution with
  logarithmic complexity,'' {\em SIAM Journal on Scientific Computing},
  vol.~35, no.~3, pp.~A1511--A1536, 2013.

\bibitem{ttgsm}
E.~Levdik and A.~A. Shcherbakov, ``{TTGSM}.''
  \url{https://github.com/CompPhysLab/ttgsm}, 2024.

\end{thebibliography}

\end{document}